\documentclass[preprint,aip,jmp,amsmath,amssymb,superscriptaddress]{revtex4-1}

\usepackage{graphicx,epstopdf}%
\usepackage{dcolumn}%
\usepackage{bm}%
\usepackage[T1]{fontenc}
\usepackage[latin9]{inputenc}
\usepackage{array}
\usepackage{rotating}
\usepackage{verbatim}
\usepackage{graphicx,epstopdf}
\usepackage{setspace}
\usepackage{url}
\usepackage[]{natbib}


\begin{document}
\title{Contemporary machine learning: a guide for practitioners in the physical sciences.}

\author{Brian K. Spears}
\email{spears9@llnl.gov}
\affiliation{Lawrence Livermore National Laboratory, P.O. Box 808, Livermore, California 94551-0808, USA}


\begin{abstract}
Machine learning is finding increasingly broad application in the physical sciences. This most often involves building a model relationship between a dependent, measurable output and an associated set of controllable, but complicated, independent inputs. We present a tutorial on current techniques in machine learning ? a jumping-off point for interested researchers to advance their work. We focus on deep neural networks with an emphasis on demystifying \emph{deep learning}.  We begin with background ideas in machine learning and some example applications from current research in plasma physics. We discuss supervised learning techniques for modeling complicated functions, beginning with familiar regression schemes, then advancing to more sophisticated deep learning methods. We also address  unsupervised learning and techniques for reducing the dimensionality of input spaces.  Along the way, we describe methods for practitioners to help ensure that their models generalize from their training data to as-yet-unseen test data.  We describe classes of tasks -- predicting scalars, handling images, fitting time-series -- and prepare the reader to choose an appropriate technique. We finally point out some limitations to modern machine learning and speculate on some ways that practitioners from the physical sciences may be particularly suited to help.

\end{abstract}

\maketitle

\section{\label{sec:intro}Diving into machine learning}
Companies today invest tens of billions of dollars every year to
develop machine learning technology, making it a ubiquitous tool for
analyzing and interpreting data.  Google and Facebook use machine
learning algorithms to serve you ads.  Amazon and Apple use machine
learning both to process spoken language and to synthesize realistic
sounding voices.  Tesla uses learning tools to develop self-driving
vehicles.  Learning techniques have also made their way into more
surprising applications: Jaguar has adopted learning tools, not to
drive their cars, but to provide mapping services that optimize
cellular service reception along the drive.  Unilever even uses
machine learning to design consumer products like shampoos.
 
Machine learning impacts more than commerce and consumer goods.  The
number of scientific applications is exploding.  In the physical
sciences, learning techniques have delivered new techniques for data
analysis and prediction, new methods for comparing simulations and
experiments, and new directions in scientific computing and computer
architecture.  Researchers from disparate disciplines have
incorporated machine learning tools across a host of applications:
fitting scattered data, fitting or recognition of vector- or
image-valued data, signal analysis, approximation of partial
differential equations, construction of smooth functions for analysis
and optimization, and much more.
 
Beyond the technical advances, nations are vying for technical
dominance in the arena, with China and the US widely perceived as
leading.  China's goal is to achieve dominance in machine learning by
2030.  Vladimir Putin announced, "Artificial intelligence is the
future ... whoever becomes the leader in this sphere will become the
ruler of the world."  In a move that scientists can expect to
influence science policy, the US House of Representatives created the
Artificial Intelligence caucus to seek science and technology input
for developing public policy\cite{caucus}.  For many reasons, then, a working
knowledge of the principles of machine learning is beneficial to
physical scientists.

Our aims are: 
\begin{enumerate}
\item to develop a foundation from which researchers can
explore machine learning,
\item to demystify and define machine learning
with an emphasis on deep learning via neural networks, 
\item to lay out
the vocabulary and essential concepts necessary to recognize the
strengths of deep learning,
\item to identify appropriate learning
techniques for specific applications, and 
\item to choose software tools
to begin research exploration.
\end{enumerate}

\section{\label{sec:definition}Machine learning: context and a definition}
Machine \emph{learning} is the application of a numerical algorithm that improves its \emph{performance} at a given \emph{task} based on \emph{experience} \cite{mitchell:defn_learn}. The \emph{task} is to predict a numerical value based on numerical input.  Mathematically, we desire a function that maps our inputs to output values, say $y = f(\bold{x})$. The \emph{experience} is the collection of input and output values, $(X,Y^*)$ where $X=\{\bold{x}_i\}$ and $Y^*=\{y^*_i\}$, with $i$ ranging over $N$ examples. These examples come to us from simulation or experimental observation. We can measure the \emph{performance} of a learning algorithm by the nearness of its predicted values, $y$, to the true target values, $y^{*}$.  In the simplest case, we might measure the performance by the squared error, $SE=\sum{(y^*_i-y_i)^2}=\sum{(y^*_i-f(\bold{x}_i))^2} $. The \emph{learning} is the improvement of the algorithm performance with exposure to additional experience or data.  Typical tasks for machine learning include classification, clustering, dimensional reduction, and regression. Our task for this tutorial will be regression -- using learning algorithms to approximate real-valued functions.

The familiar fitting methods used in the physical sciences are elementary \emph{parametric} machine learning algorithms.  The prototype is the linear least squares method.  Here, we use labeled (supervised) data, $\{(y_1,x_1),(y_2,x_2),...,(y_N,x_N)\}$, to fit a model with explicit parameters.  Examples of parametrized model functions for use with linear least squares include the familiar 
\begin{equation}
y=ax+b
\end{equation}
and the series
\begin{equation}
y=a_{0}+\sum_{k=1}^{N}(a_{k}cos(\frac{k\pi x}{L})+b_{k}sin(\frac{k\pi x}{L}))
\end{equation}
, both of which are linear in their parameters.  They clearly need not have basis functions that are linear in x.  We can relax the need for linearity in the parameters to accommodate models like
\begin{equation}
y=(ax+sin(b)x^3)^c
\end{equation}
. However, in this nonlinear case, we must appeal to nonlinear solution techniques, like the Levenberg-Marquardt procedure.  In any case, linear or nonlinear, these parametric methods require that we know a suitable basis in advance based on prior knowledge of the application at hand.

Machine learning algorithms can be extended beyond parametric techniques to \emph{non-parametric} methods.  These algorithms do not require an explicit parameterization or, in linear models, a statement of the basis.  Examples include support vector machines, decision trees, and (deep) neural networks.  In neural networks, the algorithm builds a useful representation of the data by setting a very large number of parameters.  The parameters combine many very simple functions to build up the function being approximated. It is counterintuitive that neural network techniques are considered non-parametric because they employ a large number of parameters.  But, the essential feature of non-parametric techniques, in particular neural networks, is that we need not describe a parameterization in advance based on prior knowledge.  This gives the technique valuable flexibility to fit potentially complicated and unknown details in the function to be approximated.  Avoiding the specification of a parameterization, of course, comes at a cost.  Without the constraining prior information of a parameterization, non-parametric techniques require more data for training (fitting).  This tradeoff between flexibility and data volume requirements presents a recurrent challenge as we design and execute learning algorithms.

\section{\label{sec:examples}Some motivational examples from the plasma physics community}
Contemporary advances in machine learning are being quickly incorporated into research of interest to the plasma physicists. Machine learning has been broadly investigated to help predict disruption in tokamak devices.  Disruption, the sudden loss of confinement, is both potentially damaging to the device and difficult to model and predict.  Rea and Granetz \cite{rea} have used random forest learning techniques to predict disruptions on DIII-D with high accuracy.  Here, the learning tool assigns the the observed device conditions to a category -- nondisrupted, near disruption, or far from disruption.  This categorical prediction task is called classification.  Others have developed similar predictive classification capabilities for DIII-D and JET using neural networks and support vector machines \cite{cannas:nn_disrupt, vega:disrupt}.

Researchers are also incorporating numerical techniques directly into numerical simulations.  Multiple groups have investigated using neural networks to learn closure models for hydrodynamic simulations of turbulent flow.  We consider here an illustrative proof of principle for incorporating trained neural networks directly into discretized partial differential equation (PDE) models \cite{duraisamy}.  Using the Spallart-Almaras turbulence model 
\begin{equation}
\frac{\partial \hat{\nu}}{\partial t} + u_j   \frac{\partial \hat{\nu}}{\partial x_j} = \frac{1}{\sigma}\left(\frac{\partial}{\partial x_j} \left((\nu + \hat{\nu})\frac{\partial \hat{\nu}}{\partial x_j}\right)+ c_{b2} \frac{\partial \hat{\nu}}{\partial x_i}\frac{\partial \hat{\nu}}{\partial x_i}\right) + c_{b1}(1-f_{t2})\hat{S}\hat{\nu} - \left(c_{w1}f_w-\frac{c_{b1}}{\kappa^2}f_{t2}\right)\left(\frac{\hat{\nu}}{d}\right)^2
\end{equation}
researchers trained a neural network to approximate the source terms in the model (all right hand terms excluding the diffusion term, $\frac{\partial}{\partial x_j} \left((\nu + \hat{\nu})\frac{\partial \hat{\nu}}{\partial x_j}\right)$, then performed numerical simulations showing that the model with the learned approximation reproduced the solutions of the full PDE simulations.  Similar techniques might be used in future investigations to approximate expensive physics packages with the goal of reducing computational cost.

In a final example, inertial confinement fusion (ICF) researchers used neural networks to explore high-dimensional design spaces.  The team used both random forests and deep neural networks to learn the response of an expensive radiation hydrodynamics code over a 9-dimensional parameter space.  With this learned response in hand, they navigated parameter space to find implosions that optimized a combination of high neutron yield implosion robustness.  The exercise led to the discovery of asymmetric implosions that, in simulation, provide high yield and a greater robustness to perturbations than spherical implosions.  Without the ability to search parameter space with machine learning tools, the rare, well-performing, asymmetric simulations would have been difficult, if not impossible, to find  \cite{Peterson:2017kq, humbird:djinn, Nora:coda2015}.

\section{\label{sec:fundamentals}Fundamentals of neural networks}
\begin{figure}[h]
\centering{}\includegraphics[width=0.75\columnwidth]{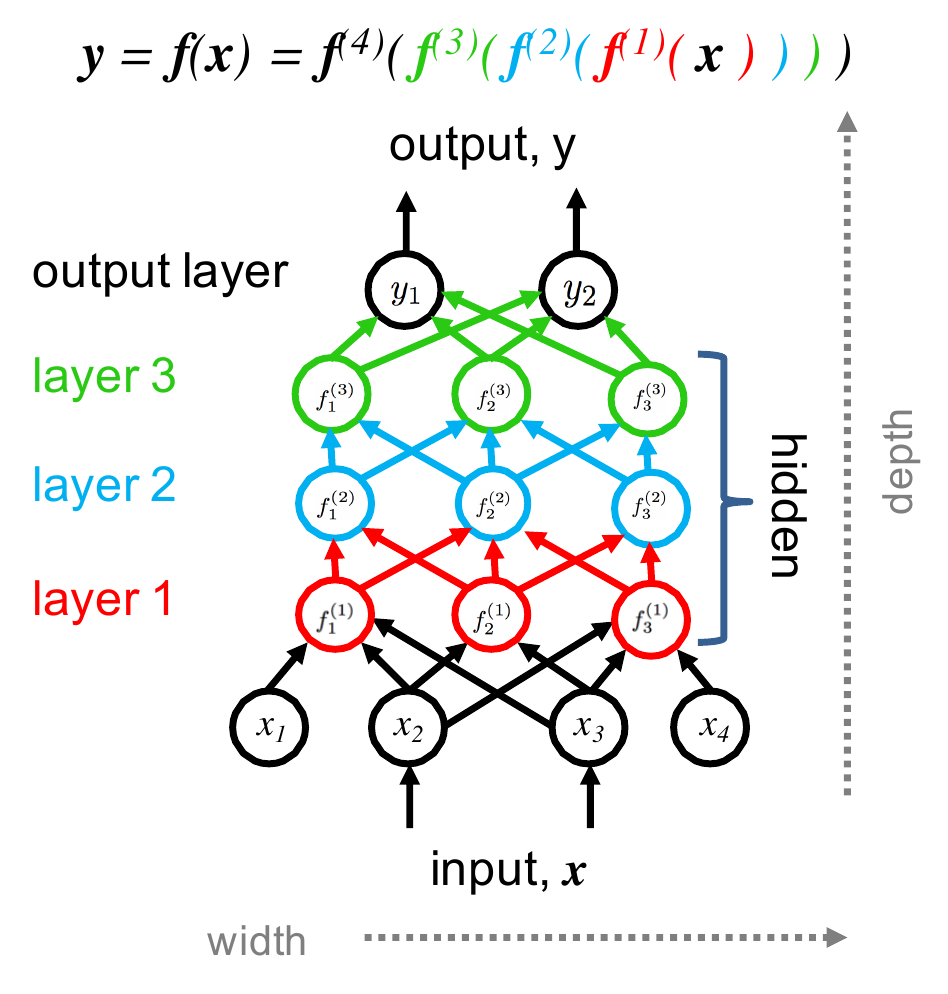}
\caption{\label{fig:network_nomenclature} Neural networks can be represented as graphs.  The edges (arrows) represent the weights and biases of linear transformations between the layers.  The circles represent the nonlinear activation functions performed by the neurons or units.  The interior (colored) layers are called hidden layers.  Network architectures are described by their depth (number of layers) and layer widths (number of units). }
\end{figure}

The most exciting growth in contemporary machine learning has come from advancements in neural network methods. A \emph{neural network} is a set of nested, nonlinear functions that can be adjusted to fit data. A neural network, then, is really a complex function of the form
\begin{equation} \mathbf{y} = \mathbf{f}(\mathbf{x}) = \mathbf{f^{(J)}( \ldots f^{(3)}(  f^{(2)}(  f^{(1)}(x)))\ldots)} \end{equation}
An example network is conveniently represented as a graph in figure \ref{fig:network_nomenclature}.  The input values, $\mathbf{x}$, experience a nonlinear transformation at each layer of the network.  The final layer, or output layer, produces the ultimate result, the predicted values, $\mathbf{y}$.  Intermediate layers are called \emph{hidden layers} since their inputs and outputs are buried within the network.  Each of these layers is composed of a unit, or neuron. A network layer can be described by its \emph{width}, or the number of units in the layer.  The network can also be described by the total number of layers, or the \emph{depth}.  Many-layer networks, or deep neural networks, frequently outperform shallow ones supporting the heavy interest in \emph{deep learning}.

\begin{figure}[h]
\centering{}\includegraphics[width=0.75\columnwidth]{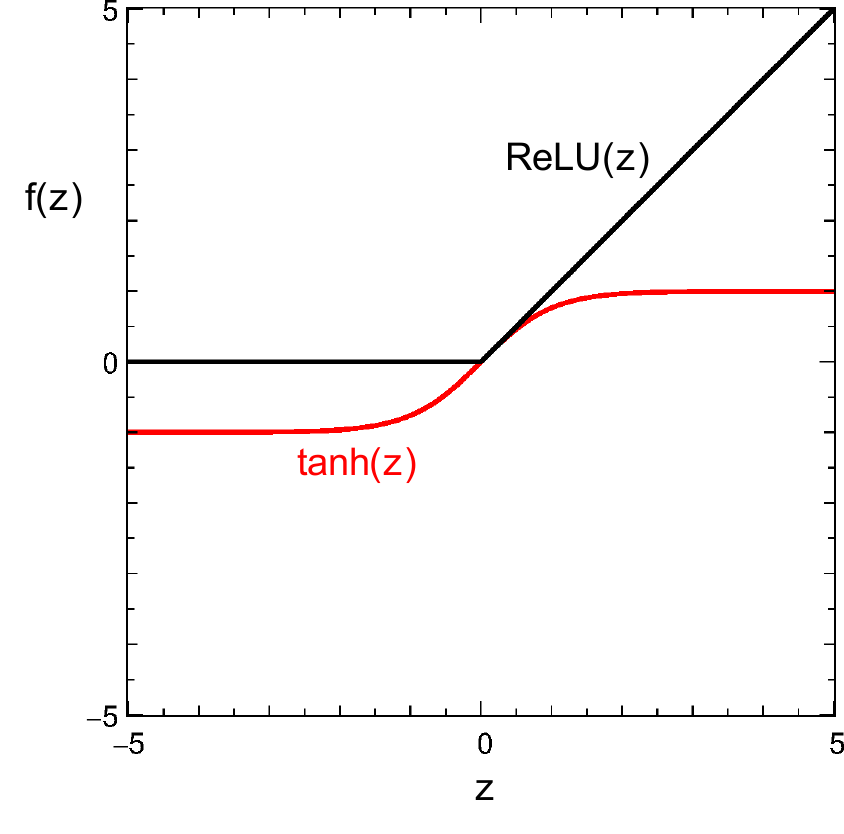}
\caption{\label{fig:activation} Activation functions are the nonlinear transformation performed by each neuron in a network.  Historically, neural networks have used sigmoidal functions that saturate, like $\tanh(z)$.  Modern networks achieve improved performance using the ReLU$(z)$ function to rectify shortcomings of sigmoidal functions.}
\end{figure}

Each neuron in a layer operates on a linear combination of the values from a previous layer such that a subsequent layer accepts values $\mathbf{z}$ constructed from the prior layer outputs, $\mathbf{x}$, as 
\begin{equation}\mathbf{z=Wx+b}\end{equation}
The elements in the tensor, $\mathbf{W}$, are known as the \emph{weights} and in vector, $\mathbf{b}$, as the \emph{biases}. The weights and biases are the (many) free parameters to be chosen to approximate the relationship between inputs and outputs in a set of data to be fitted. The nonlinear operation performed by each unit is known as the \emph{activation function}.  We show candidate activation functions in figure \ref{fig:activation}. Historically, the activation function was sigmoidal, like $\tanh(z)$.  Current practice relies heavily on the rectified linear unit, or $ReLU(z)$.  This piecewise linear, but globally nonlinear, often yields much better results than sigmoidal functions.  This is mainly attributed to the saturation behavior of sigmoidal functions that can lead to shallow gradients that slow learning. Taking advantage of the linear combinations between layers and choosing ReLU as the activation function, our example neural network becomes
\begin{equation} \mathbf{y} = \mathbf{f}(\mathbf{x}) = \mathbf{f^{(4)}(b^{(4)}+W^{(4)}  f^{(3)}(  b^{(3)}+W^{(3)}f^{(2)}(b^{(2)}+W^{(2)}f^{(1)}(b^{(1)}+W^{(1)}x))))} \end{equation} where the $f(z) = ReLU(z) = max\{0,z\}$

\begin{figure}[h]
\centering{}\includegraphics[width=0.75\columnwidth]{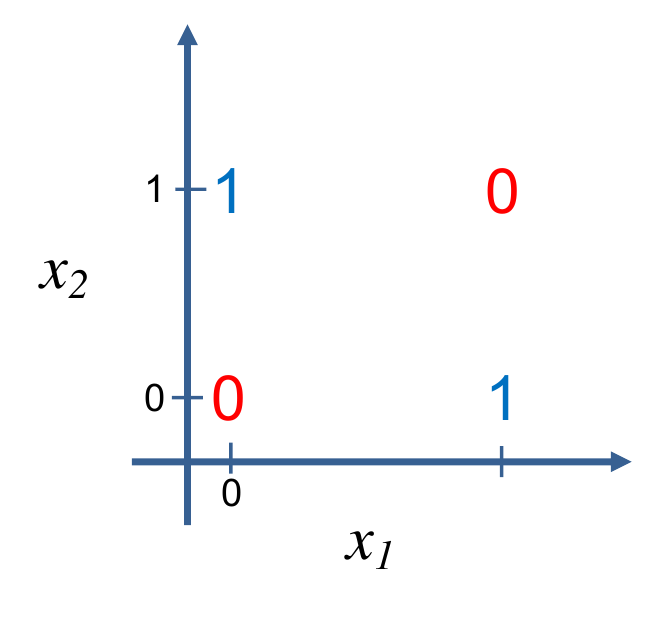}
\caption{\label{fig:xor_values} The exclusive-or (XOR) function is a nonlinear function that returns 0 when its two binary inputs are both 0 or both 1.  It returns 1 when its binary inputs are different. The XOR cannot be represented by a linear network or a two-layer network.  A deeper, 3-layer network with a nonlinear activation function can very easily represent the XOR.}
\end{figure}

To cement our understanding of the basics of neural networks, we turn to an instructive, analytical example.  We will develop a small network to learn the exclusive or function, XOR.  The XOR, represented in figure \ref{fig:xor_values}, accepts independent variables $x_1$ and $x_2$.  When both input values are $1$ or both values are $0$, XOR returns 0.  When $x_1$ and $x_2$ are different from each other, XOR returns $1$.  Using our language from section \ref{sec:definition}, our task is to regress on the experience $X=\{(0,0),(0,1),(1,0),(1,1)\}$ with supervised labels $Y=\{0,1,1,0\}$, respectively.

The example is not only interesting because we can write down the solution without appealing to extensive numerics, but also because it is of historical importance.  Critics of neural networks in the 1980's (check dates) noted that the XOR problem could not be solved with a 2-layer network.  This lead critics to generalize, wrongly, that deep neural networks might also fail to handle essential nonlinearities in learning tasks.  It is now well known that deep networks are exceptionally powerful for handling richly nonlinear tasks.  

\begin{figure}[h]
\centering{}\includegraphics[width=0.35\columnwidth]{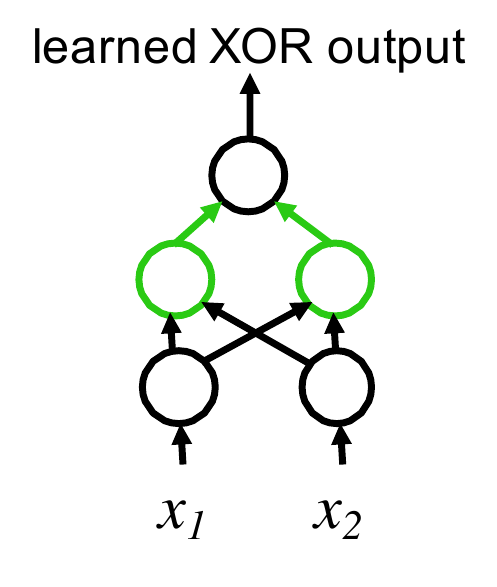}
\caption{\label{fig:xor_net} A shallow, narrow neural network architecture is sufficient to represent the XOR function, provided the activation function is chosen appropriately.  While a linear activation function (and therefore linear network) fails, a ReLU$(z)=\max\{0,z\}$ function successfully describes the XOR as $y=\max\{0,x_1+x_2\}-2\ \max\{0,x_1+x_2-1\}$. }
\end{figure}

We proceed here to show that a 3-layer network (figure \ref{fig:xor_net}) succeeds at the XOR task. Our treatment is a modification of an example from the excellent book, \emph{Deep Learning} \cite{Goodfellow:deep_learning}. We take the opportunity to emphasize the importance of our choice of activation function to the network performance.  We will experiment with two activation functions: a linear function (bad choice) and the ReLU (good choice).  We begin with the linear activation function.  At this point, we have specified our network architecture (figure \ref{fig:xor_net}) and our activation function (linear).  We next choose the cost function we use to measure the nearness of our predicted values to the true XOR values.  For simplicity, we choose mean squared error such that
\begin{equation} J(\pmb{\theta})=\sum_{x\in X}(XOR(\mathbf{x})-f(\mathbf{x};\pmb{\theta}))^2\end{equation}
Our network approximation is very simple:
\begin{equation} f(\mathbf{x};\pmb{\theta})=f(\mathbf{x};\mathbf{w},b)=\mathbf{w}\cdot\mathbf{x}+b\end{equation}
Inserting into the cost function, we recover the normal equations for linear least squares.  The solution is $\pmb{w=0}$ and $b=\frac{1}{2}$.  This constant solution is not at all what we want.

Let us now explore the same procedure -- same network, same loss function, but this time choosing ReLU for the activation function.  Calling the input, $\pmb{x}$, the hidden layer output, $\pmb{h}$, and the final scalar output, $y$, we have

\begin{equation} \pmb{h} = g(\pmb{W}\pmb{x} + \pmb{c})\end{equation}
as the transform from input layer to hidden layer and
\begin{equation}y=\pmb{w\cdot h}+b\end{equation}
as the transform from hidden layer to final output. Combining the transformations, we have (summing on repeated indices)
\begin{align}
y &=w_i \ g(W_{ji} x_j + c_i) +b\\
  &=w_i \max\{0,W_{ji} x_j + c_i\} +b
\end{align}
 
 We now have a neural network, albeit a simple one.  What remains is to select the indexed constants.  We could try to learn these constants using the training experience and an optimization algorithm like gradient descent, which we describe next.  For now, we simply select the nine numbers needed to exactly reproduce the XOR behavior.  This leads to a completely specified network
 \begin{equation}y=\max\{0,x_1+x_2\}-2\ \max\{0,x_1+x_2-1\}\end{equation}
 which by inspection can be seen to give the desired answers.  This simple example has served two purposes for us.  It has made concrete what a neural network is, but has it also highlighted the importance of the proper activation function.  We must exercise caution when choosing this function in practical applications, too. 

Of course, deep learning is interesting because it scales well to enormously difficult research tasks.  For these research tasks, we need a numerical method for selecting the optimal parameters when we cannot surmise them by inspection.  In these cases, we seek a technique for minimizing the cost function.  The standard example process is as follows:
\begin{enumerate}
\item compute current estimates of output, $y=f(x;\pmb{\theta})$
\item measure the difference between current estimates and true training data using the loss function,  $J(\pmb{\theta})=\sum_{x\in X}(y^*(x)-f(x;\pmb{\theta}))^2$
\item compute the gradient of the loss function with respect to the parameters, $\theta$, using \emph{backpropagation}
\item choose new parameters that most reduce the loss function using \emph{gradient descent}
\end{enumerate}

\emph{Backpropagation} is an efficient algorithm to compute the gradient of the loss function with respect to the parameters, $\pmb{\theta}$.  Because the training data is independent of the choice of $\pmb{\theta}$, this is really an algorithm for finding the gradient of the network itself
\begin{equation}\nabla_{\pmb{\theta}}y= \nabla_{\pmb{\theta}}f(x;\pmb{\theta})\end{equation}. The algorithm specifies the order of differentiation operations following the chain rule so that repeatedly used derivatives are stored in memory rather than recomputed.  This accelerates the computation, instead burdening memory, which is desirable for most applications.  

With the gradient in hand, a gradient descent algorithm can be used to update parameters according to a rule like
\begin{equation}\pmb{\theta}_{new} = \pmb{\theta}_{old} + \epsilon \nabla_{\pmb{\theta}}f(x;\pmb{\theta})\end{equation}. The parameter $\epsilon$ is commonly called the learning rate.  We must set the learning rate with care.  The nonlinear nature of deep neural networks typically introduces many local minima.  Setting the learning rate too small can trap the gradient descent in a sub-optimal local minimum.  Setting it too large can allow large leaps that skip regions of desirable behavior.  There are also alternative parameter optimization techniques, including ones with variable learning rates and Newton-style schemes. 

\section{\label{sec:toy}A numerical starting point}
We now turn to a simple numerical example to help develop the numerical tools required for application of deep neural networks.  Our task will be to develop an approximate function for the simple, nonlinear relationship $y=x_1^2 + x_2^2$.  We will use the open-source Python package \texttt{scikit-learn} \cite{scikit-learn} to help readers begin.

\begin{verbatim}
from sklearn.neural_network import MLPRegressor
x1, x2 = mgrid[-1:1:200j, -1:1:200j]
v1     = ravel(x1)
v2     = ravel(x2)
Y      = v1**2 + v2**2
X  = stack((v1,v2),axis=1)
nn = neural_network.MLPRegressor()
nn.fit(X,Y)
yptrain = nn.predict(X)
\end{verbatim}

\begin{figure}[h]
\centering{}\includegraphics[width=0.75\columnwidth]{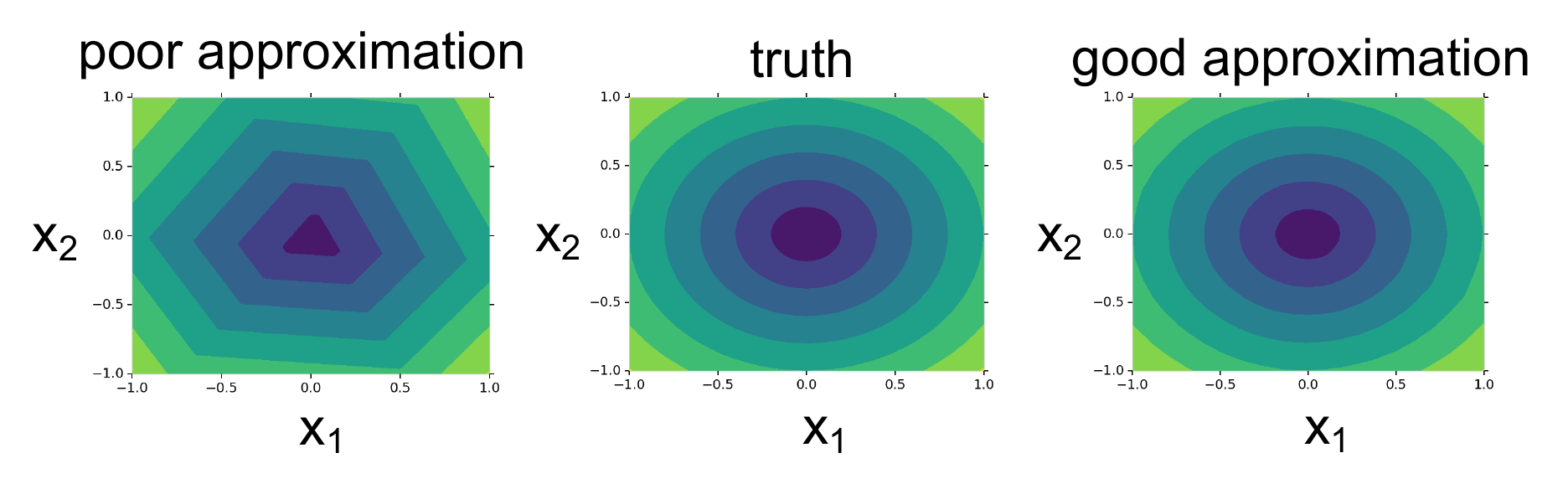}\caption{\label{fig:toy_problem} The multi-layer perceptron, or deep neural network, tool in \texttt{scikit-learn} \cite{scikit-learn} can readily represent the simple example function of section \ref{sec:toy}.  With badly chosen hyperparameters (network architecture, regularization strength, etc.), the network is a poor approximation (left panel) of the true function values (central panel).  With well-chosen hyperparameters, the network is a good approximation (right panel) of the truth.
}
\end{figure}

Here, the class \texttt{MLPRegressor} (a \textbf{M}ulti\textbf{L}ayer \textbf{P}erceptron, or deep neural network), returns a neural network object.  The method \texttt{fit()} performs backpropagation and gradient descent using the training data \texttt{X,Y}.  Then, the method \texttt{predict()} evaluates the trained neural network at all locations in the data \texttt{X}.  Software tools like \texttt{MLPRegressor} are helpful because they can be implemented with relative ease.  However, even simple deep learning techniques are powerful and flexible.  They require the user to set or accept defaults for multiple parameters, for example hidden layer sizes, learning rate, activation function, etc.  The efficient choice for these requires knowledge of the underlying numerics and often some experimentation.  We show in figure \ref{fig:toy_problem} the true function and neural neural network approximations made with both poor and good choices of parameters.

\section{\label{sec:fitting}Examining the quality of your learned model}
\begin{figure}[h]
\centering{}\includegraphics[width=0.75\columnwidth]{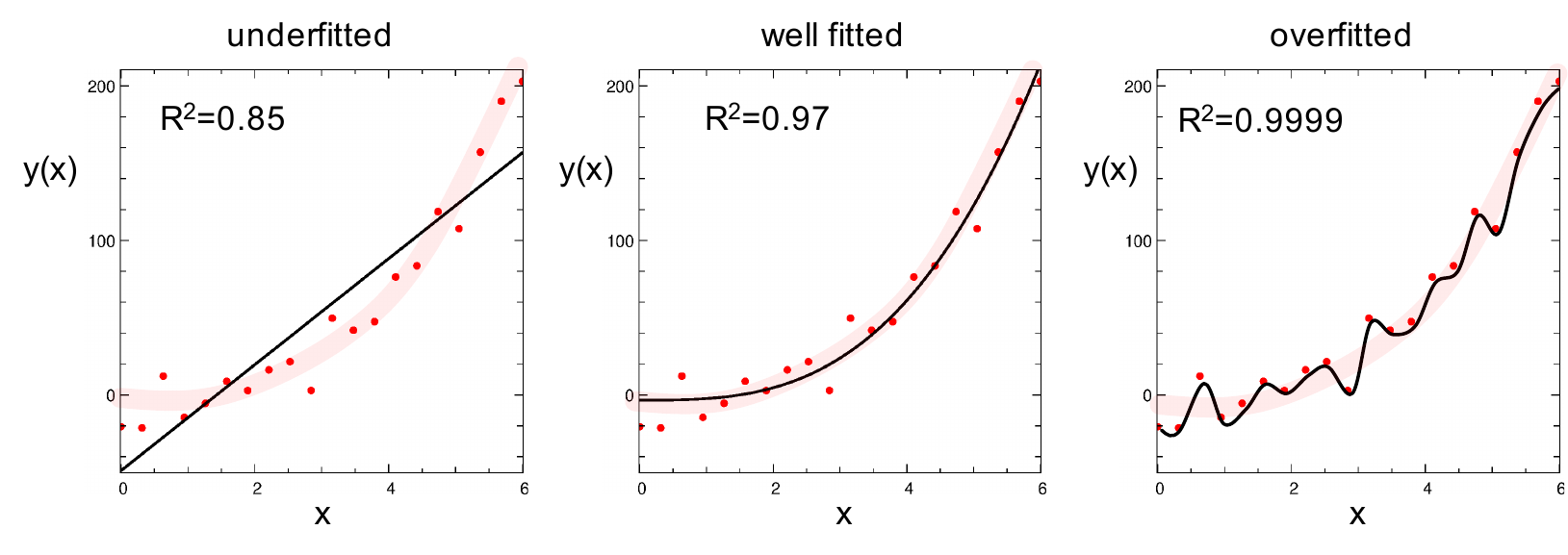}\caption{\label{fig:fit_quality} Goodness of fit must be judged based on how well the model performs on both training data and unseen test data.  The sequence of models represents increasing nearness to the training data (dots).  As measured by $R^2$, the model goodness of fit increases as the model capacity increases from left to right.  However, given the quadratically distributed training data, the right panel is overfitted -- it will fair poorly on new data that is not part of the training set.}
\end{figure}

This raises a key question: what does it mean for a learned model to be good?  We can begin by defining a scalar measure for goodness of fit like the $R^2$ value
\begin{equation} R^2 = 1-\sum_{i=1}^{n}\frac{(t_i-p_i)^2}{(t_i-E[t])^2}\end{equation} where $t_i$ are the true training values, $p_i$ are the predicted values, and $E[t]$ is the expectation value of the multiple $t_i$.  As the $p_i$ approach the $t_i$, $R^2$ tends to unity.  However, it is not sufficient for the model to achieve a high $R^2$ value on the training data.  We show a set of three model fits in \ref{fig:fit_quality}.  The best model achieves an $R^2$ of $0.97$ and is intuitively what we mean by a good fit.  We call this a well fitted model. The model with low $R^2$ is a bad fit and uses a model that is too simple to explain the data.  We call this failure to match the training data \emph{underfitting}. The model with $R^2>0.99$ has a good fitness metric, but is clearly overly complicated for the data.  We call this behavior \emph{overfitting}.  All of our fitness assessments have been made on the same data that we used to train our models.  We call this an assessment of \emph{training error}.

With simple univariate data, it is sometimes possible to identify underfitting or overfitting by plotting both the model and the training data against the independent variable.  However, we need to be more sophisticated with the high-dimensional data typical to deep learning applications.  To do so, we introduce the notion of generalization to our model.  We demand not only that the fitted model get the right answer for data that was used in training, but also that it generalize -- that it get the right answer for data that was \textbf{not} used in the training.  We can compute a \emph{generalization error}, or \emph{test error}, using the same $R^2$ function to assess data not used in training.  This data might be subset of the available training data that was intentionally held out to test generalization, or it might be new data collected after training. The concept of testing both training error and generalization error is called \emph{cross validation}.  

\begin{figure}[h]
\centering{}\includegraphics[width=0.75\columnwidth]{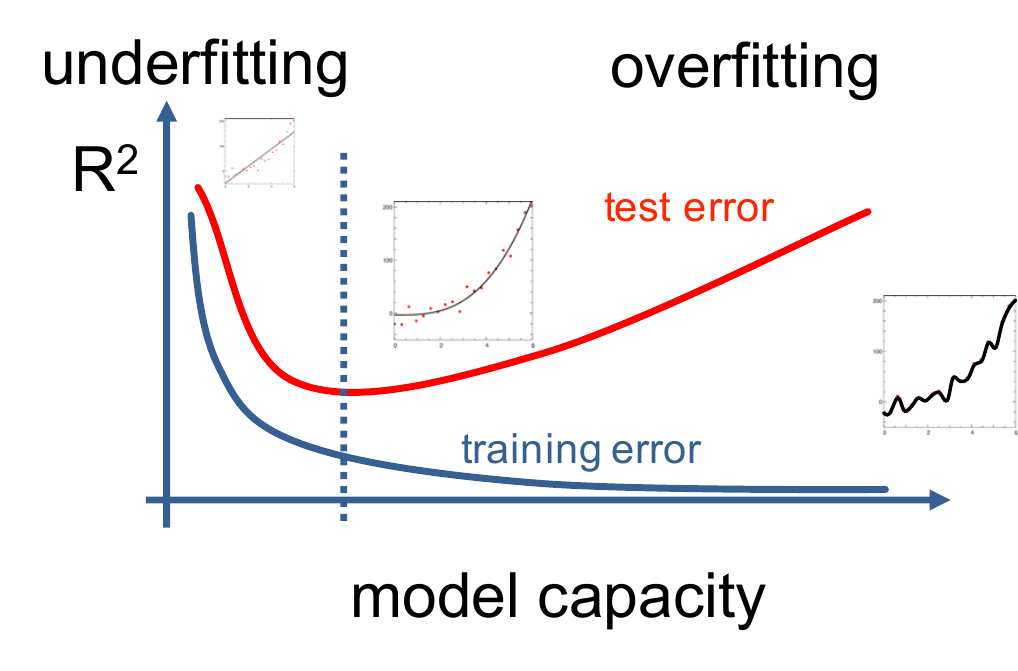}\caption{\label{fig:model_capacity} Models generally exhibit reduced training error as we increase capacity.  However, the test error eventually increases with capacity as we begin to overfit. We can adjust model capacity to optimize fit quality, minimizing the difference between test and training error (dashed line).}
\end{figure}

While developing a reliable trained model, we usually adjust the model capacity, or the flexibility with which it can accommodate the data.  We can add capacity by introducing additional neurons or layers, for example.  We can remove capacity by adding a cost function penalty (regularization) for regions of parameter space that produce undesirable models.  As we increase model capacity the test and training errors typically evolve as shown in figure \ref{fig:model_capacity}.  The training error falls to low values as the model "connects the dots," or directly interpolates the data.  However, the test error reaches a minimum before rebounding.  As the model becomes overly complicated, it begins to fail to predict unseen test data.  Our models are underfitted if they have high training error.  Once we have increased the model capacity to reduce training error, we turn to the training error.  Models with low training error, but high test error, are overfitted.  For intermediate capacities, the model is said to be well fitted.  It may be that even in the well-fitting regime, we find the test error unacceptably high.  In this case, we may be forced to collect more training data to improve the fit.  This is usually an expensive or time-consuming proposition.

\section{\label{sec:deep}The strengths of deep learning solutions}
In principle, neural networks can offer perfect approximations to functions.  This notion is described formally and theoretically in work on universal approximation.  Multiple authors have shown that any sufficiently smooth function can be represented by a 3-layer neural network \cite{cybenko,hornik}.  To be capable of universal approximation, the network must have a nonlinear (squashing) activation function.  While such a network can be proven to exist, it may not be very useful.  First, the network may need to be arbitrarily wide, making it impossible to develop enough data for training.  Second, the even the existence of a finite network says nothing about whether the network can be trained.  Much prior work has been done using sigmoidal activation functions.  Though they meet the nonlinearity requirements for universal representation, they also saturate at extreme input values.  This saturation often leads to shallow gradients in the cost function which greatly slow the training process (see section \ref{sec:fundamentals}).  The cost function can sometimes be chosen to rectify these shallow gradients, but not always.

The revolution in contemporary deep learning has been based on successful repairs to the shortcomings of historical networks. A key advance is the now-routine use of nonlinear activation functions that don't saturate (e.g., ReLU).  Networks also commonly use cost functions that are engineered to interact well with the selected activation function (e.g., cross entropy).  Perhaps the most useful advance is the recognition that deep networks routinely outperform shallow ones.  Deep networks typically require fewer total units for the same task and produce improved generalization error.  These features couple well with a host of other advancements: the development of backpropagation for efficient gradient computation, the arrival of "big data" for training large networks, modern computer architectures and processor development (e.g., the general purpose graphics processing unit (GPGPU)), and neural network architectures that can exploit structures in the training data. Taken together, these advances have propelled the explosion of progress in deep learning. 

\begin{figure}[h]
\centering{}\includegraphics[width=0.75\columnwidth]{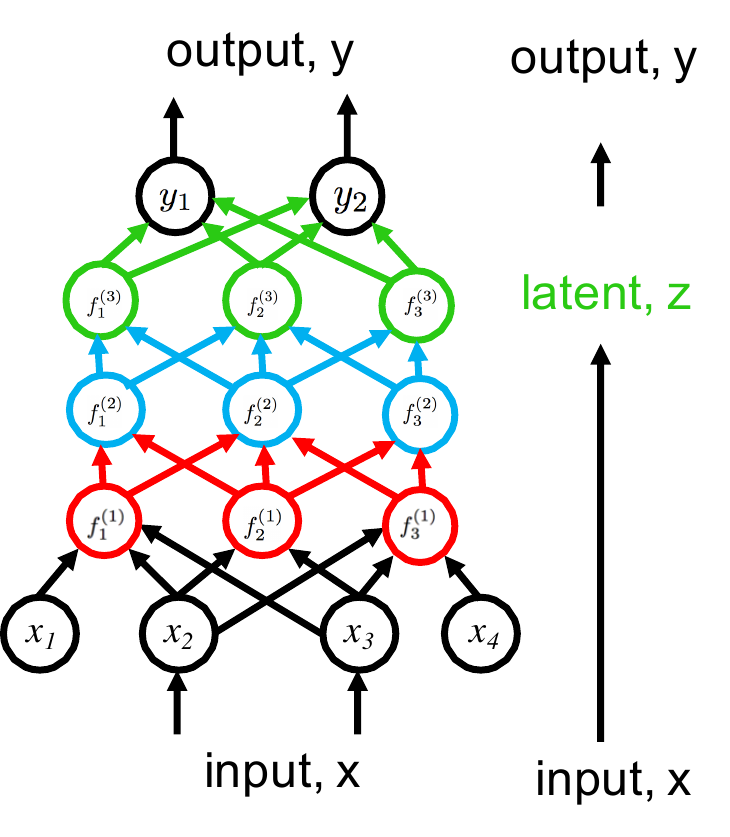}\caption{\label{fig:network_latent} Deep neural networks develop efficient representations of the input data using intermediate, latent variables.  These variables arise from the sequence of nonlinear transformations produced by the hidden layers.  The latent variables form a set of features from which it is easy to map to the desired output}
\end{figure}

The distinguishing feature of deep learning techniques is their ability to build very efficient representations of the training data.  Deep networks use the many hidden layers to develop an intermediate representation of the data called a \emph{latent space} (see figure \ref{fig:network_latent}). This latent space is essentially a nonlinear coordinate transformation. We can think of this as something like a basis for expressing the training data.  Deep neural networks rely on these effective latent spaces to capture fine details in the mapping from input to output.

\begin{figure}[h]
\centering{}\includegraphics[width=0.75\columnwidth]{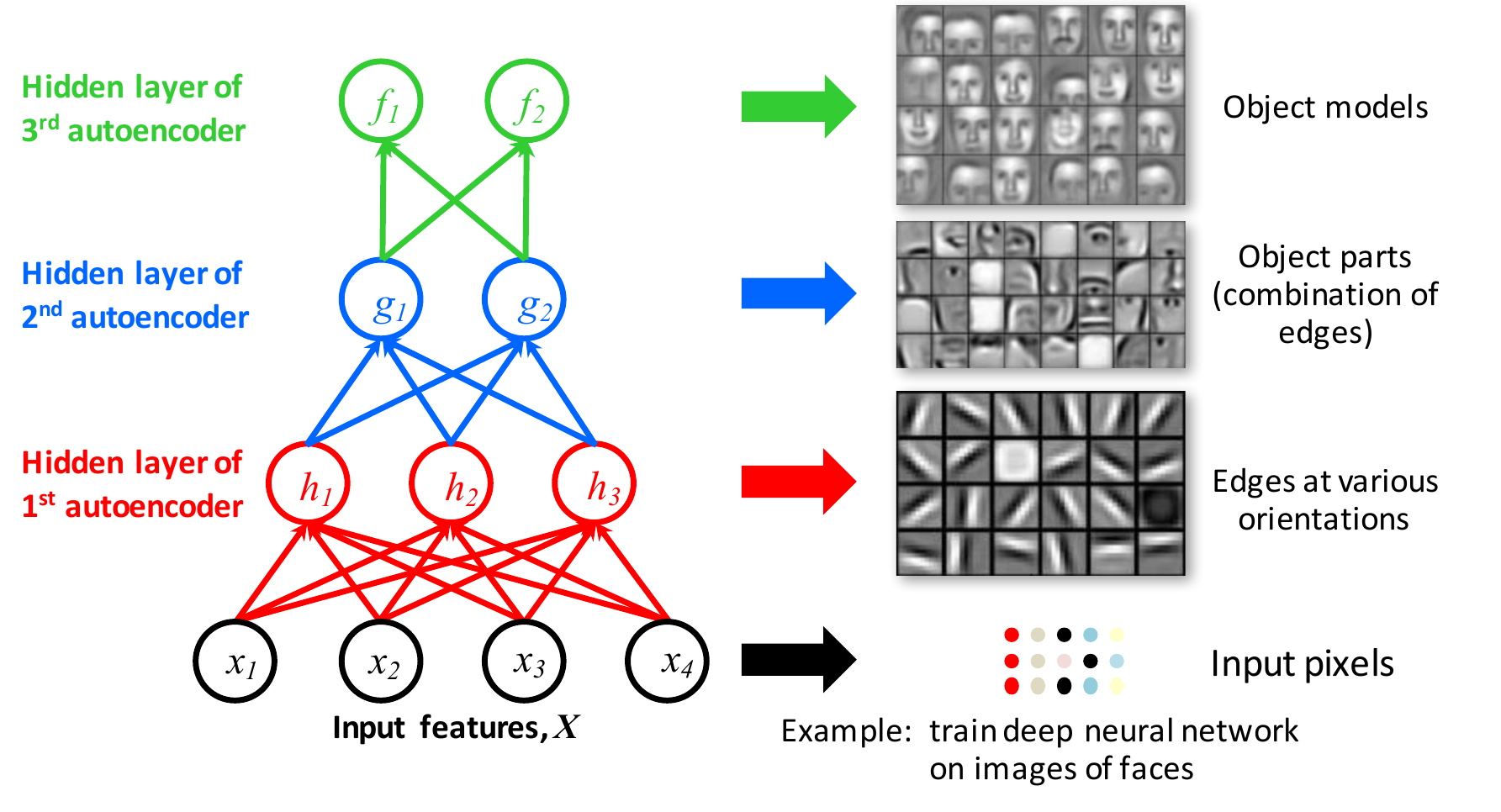}\caption{\label{fig:latent_example} This facial recognition example (modified from Honglak Lee et al.\cite{lee_honglak:latent}) shows the development of interpretable features with each hidden layer.  Eventually, the network develops a descriptive latent space of model objects from which new faces can be predicted.}
\end{figure}

\begin{figure}[h]
\centering{}\includegraphics[width=0.75\columnwidth]{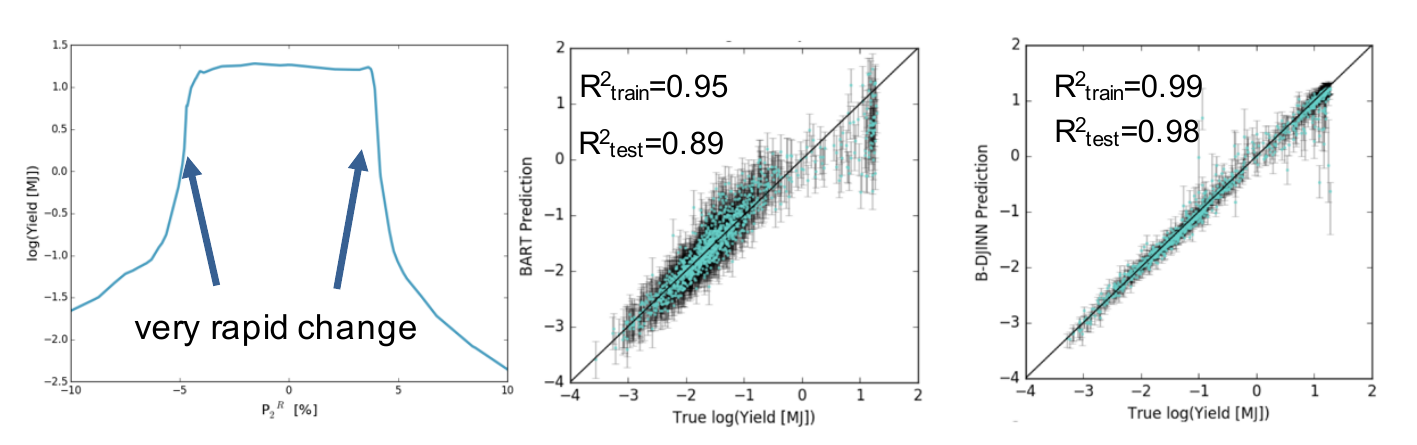}\caption{\label{fig:deep_strength} Deep neural networks excel at capturing the strong nonlinearities of ICF physics.  We show in the left panel the strong change in energy yield with a parameter, $P_2$, that controls implosion shape.  Traditional machine learning using Bayesian Additive Regression Trees (BART) fails to capture the rapid change (see poor predictions for $\log$(yield) between -1 and 1.  Deep learning techniques, like DJINN \cite{humbrid:djinn}, use well-developed latent spaces to capture the strong nonlinearity.}
\end{figure}

The notion of the latent space and the associated sequential transformations in hidden layers is beautifully described in an example by Honglak Lee et al. \cite{lee_honglak:latent} which we partly reproduce in figure \ref{fig:latent_example}.  At each layer of a neural network developed for facial recognition, we can see the structure of the latent space develop.  Each layer develops more resolving power, leading to features that can be interpreted and can also be combined to produce a desired output.  Deep neural networks like this work very well for the strong nonlinearities that can characterize plasma physics problems. We show an ICF example in figure \ref{fig:deep_strength}.  The task in this example is to reproduce the very rapid change in total neutron yield for an ICF implosion experiencing strong degradations.  While a more traditional learning model, like Bayesian additive regression trees (BART), achieves moderate training error, it generalizes rather poorly.  A deep neural network tool (called DJINN), captures the nonlinearities and generalizes well.  The network built here is considerably more sophisticated than the demonstration network in \ref{sec:toy}.  It was developed using the software package TensorFlow (\url{www.tensorflow.org}), which is specifically designed for complicated networks and large scale data. 

\section{\label{sec:tailoring}Tailoring deep networks to your application}
Deep neural networks and their efficient latent spaces are flexible tools that can be applied to many tasks.  However, the network can and should be specialized to the task.  We cover here a few common tasks that occur in physical science problems and the specialized networks that best handle them.

\subsection{\label{sub:autoencoders}Autoencoders for dimensional reduction}
We touch first on autoencoders. Autoencoders are networks composed of two consecutive pieces, an encoder and a decoder.  The encoder transforms the network input data to a more efficient representation in latent space.  The decoder reverses the the transformation, restoring the network input from the latent space representation.  Because the network maps input back to input, this is an \emph{unsupervised} learning technique.  In our initial definition of learning, supervised training used paired input and output sets, $(X,Y)$.  Here, we use only a single set as network input, say $Y$. 

\begin{figure}[h]
\centering{}\includegraphics[width=0.75\columnwidth]{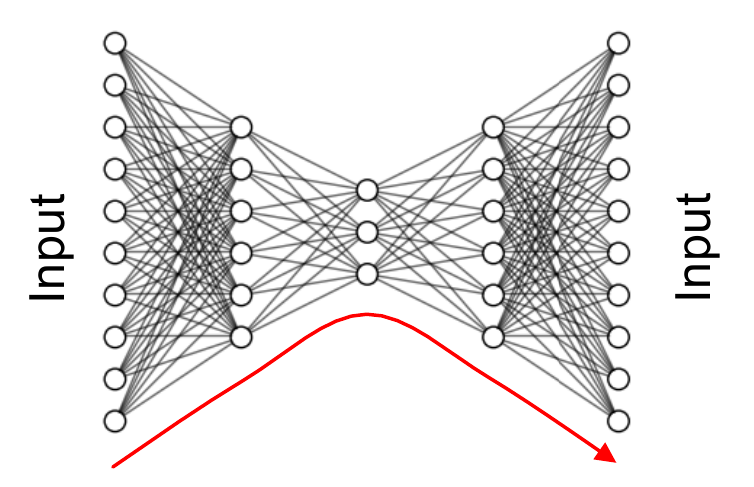}\caption{\label{fig:auto_network} Autoencoders map their input data back to itself through a reduced bottleneck.  This forces the network to develop a low-dimensional intermediate latent representation while still faithfully reproducing the input.}
\end{figure}

Autoencoders have a characteristic bottleneck structure (see figure \ref{fig:auto_network}) to compress information into a lower-dimensional latent space.  The overarching goal is usually to develop a descriptive latent representation of the data while maintaining good fidelity following decoding.  These networks can be used to reduce the dimensionality of data analogous to a principal components method.  This type of dimensional reduction is useful in data analysis and learning tasks.  Reducing the number of dimensions can reduce the volume of data needed to train models and perform analyses.  As an example, we show a dimensionally reduced autoencoder representation of x-ray spectral data \cite{humbird:spectra}.  The network successfully reduces the number variables necessary to describe the spectrum from 250 to 8.  This reduction is close to that achieved by a parameterized physics model created with expert knowledge \cite{oxford:mix_spectra}.  However, because it is a non-parameteric technique, the autoencoder did not require the parametric description of the model.

\begin{figure}[h]
\centering{}\includegraphics[width=0.75\columnwidth]{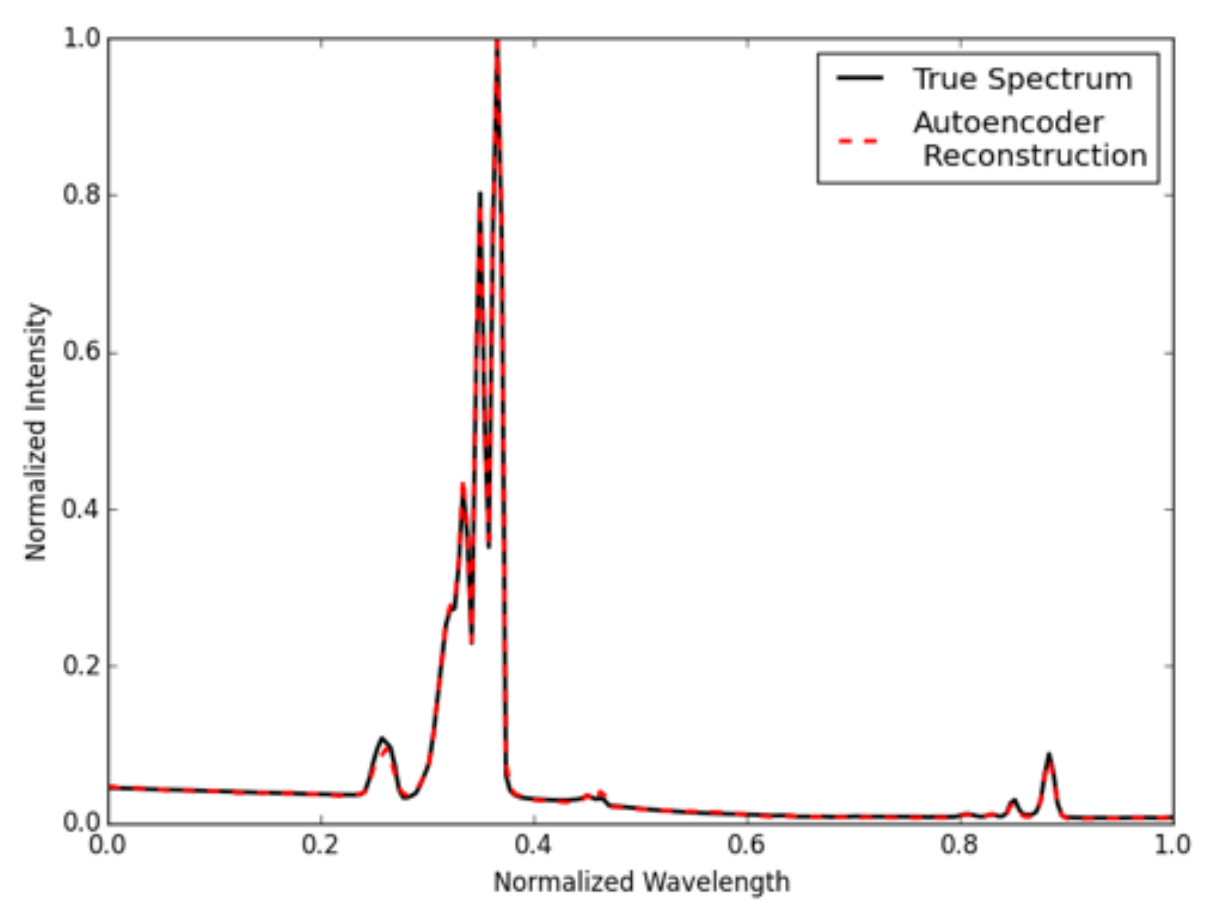}\caption{\label{fig:auto_example} Autoencoders can be designed to reduce the dimensionality of data.  We show a low-dimensional reconstruction of the detailed features of a plasma emission spectrum using an autoencoder.  The 8-parameter autoencoder model compares well with a 10-parameter, expert-designed parameteric model \cite{oxford:mix_spectra}.}
\end{figure}

\subsection{\label{sub:convolutional}Convolutional networks for arrayed data}
Neural networks can be specialized and simplified to account for structure and correlation in the training data.  We discuss now modifications that may be suitable for treating array data, whether image data or fixed-length vector data.  Here, the neighboring pixels values are often correlated.  Well-designed networks can encode these relationships in the structure of the model.  The neural network of choice is typically a \emph{convolutional network}.  

\begin{figure}[h]
\centering{}\includegraphics[width=0.75\columnwidth]{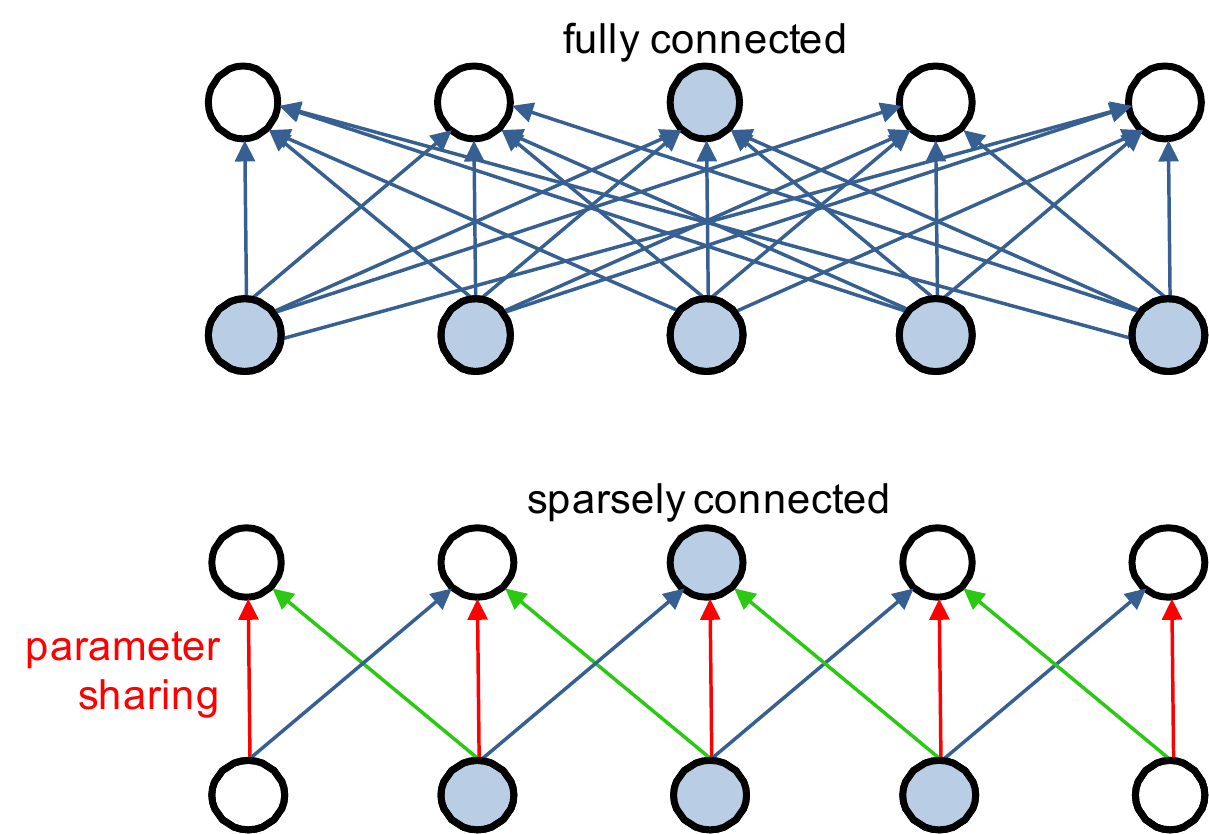}\caption{\label{fig:connectivity} Network architecture can be tailored to the data and task.  In fully connected networks, each neuron is connected to all neurons in the previous layer.   In sparsely connected networks, a neuron may be connected ton only a subset of the neurons in the preceding layer (reduced receptive field). Parameters may also be shared across edges -- all similarly colored edges have the same weight. Sparse connectivity reduces the number of parameters and the data volume required for training.}
\end{figure}

To start, we recognize that the network architecture determines the relationships between the input layer and other neurons.  While the most general neural network is fully connected, with each neuron providing input to \emph{every} neuron in the next layer (see figure \ref{fig:connectivity}), the network need not be fully connected.  In fact, the data to be learned may not support the many connections in a fully connected network.  Furthermore, we may want to modify the network to reduce its size, accelerate training, or improve its accuracy.  For example, a pixel in the center of an image likely depends on its nearest neighbors, but it is probably much less affected by the corners of the image. We might then employ \emph{sparse connectivity}. A sparse network reduces the number of connections, allowing a neuron to feed only a few near neighbors in the subsequent layer.  This reduces the number of weights and biases to be trained, consequently reducing the data required for training.  Sparse connections also change the \emph{receptive field} for each neuron.  In a fully connected network, the activation for a particular neuron depends on the inputs from all neurons in the previous layer. The receptive field for the neuron is the entire previous layer. In the sparsely connected example, the receptive field is reduced to only three nearby neurons in the preceding layer.  This reduces the impact of far-field information on local neuron values, and may better reflect the underlying data, as in our central pixel example. 

The network can be further modified to reduce the number of free parameters using parameter sharing.  In this scheme, the the weights on edges connecting neurons in the same relative position are the same.  We represent this shared weighting with color in figure \ref{fig:connectivity}.  Each directly downstream neuron has the same weight; edges on major diagonals likewise share values.  This is especially sensible if pixel is dependent on its neighbors in the same way, regardless of pixel position in the array -- a good assumption for most scientific images. 

Ultimately, to accommodate the correlations in array data, we replace the matrix multiplication in the neural network with convolution over a kernel.  This not only reduces the data required to train thanks to sparse connections and parameter sharing, but it greatly reduces the number of numerical operations needed in training.  Convolution also builds in a degree of invariance to small displacements, simplifying registration requirements in the analysis process.  In practice, convolutional neural networks have been responsible for a dramatic improvement in deep learning for image processing.  Each year, learning experts compete to develop image recognition tools using an open source image data set called ImageNet \cite{imagenet} (\url{http://www.image-net.org/}).  Until 2012, the winning error rate was about 25\%, falling a percent or two per year.  The introduction of convolutional networks in 2012 brought a 10\% reduction, and top error rates are now routinely in the low single digits.  We note here that at the same time that convolutional networks were being introduced, training on graphics processing units (GPUs) arrived, leading to computational hardware developments to support the software advancements. 

\subsection{\label{sub:transfer}Transfer learning for sparse data}
\begin{figure}[h]
\centering{}\includegraphics[width=0.75\columnwidth]{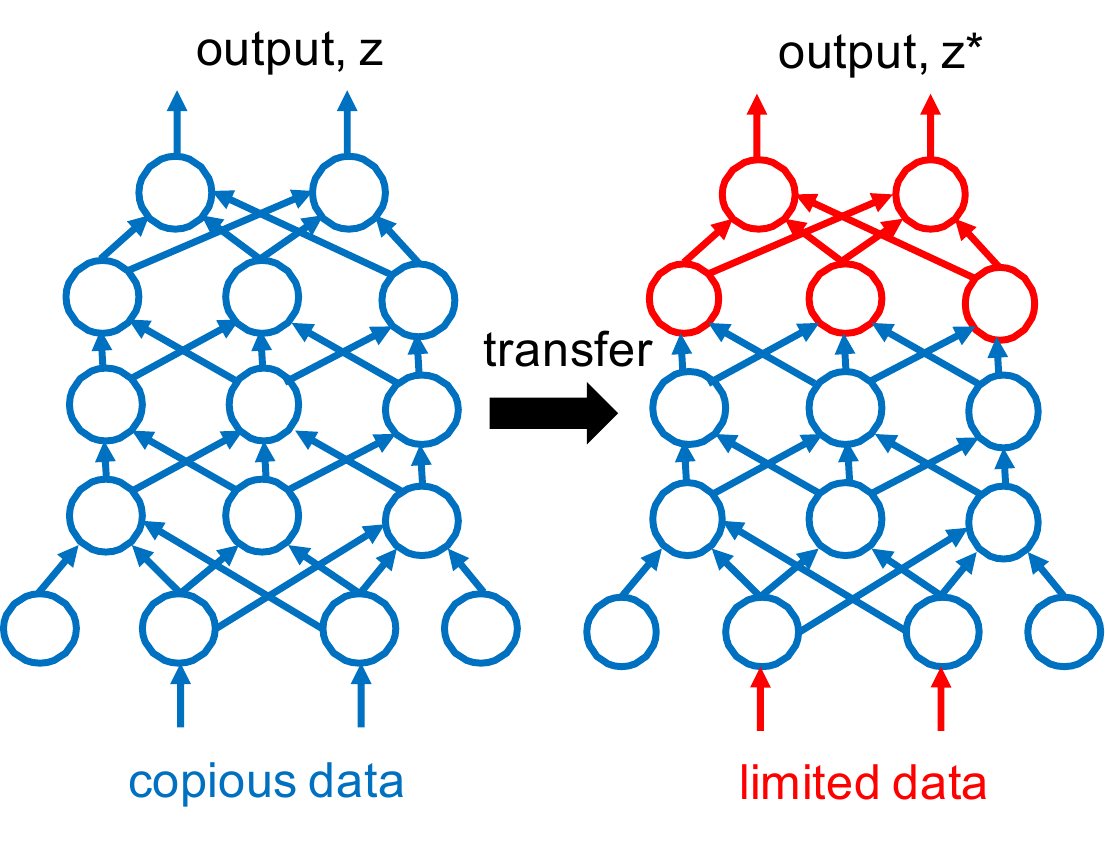}\caption{\label{fig:transfer} Transfer learning allows us to train an entire network on high volumes of readily available data (left network).  Then, a limited set of weights in the the network, say those in the final (red) layers (right network), can be re-trained on more precious, limited data. This allows a network to develop rough prediction capability on the large data set, while refining that prediction on the smaller, more specific data set. 
}
\end{figure}

\begin{figure}[h]
\centering{}\includegraphics[width=0.75\columnwidth]{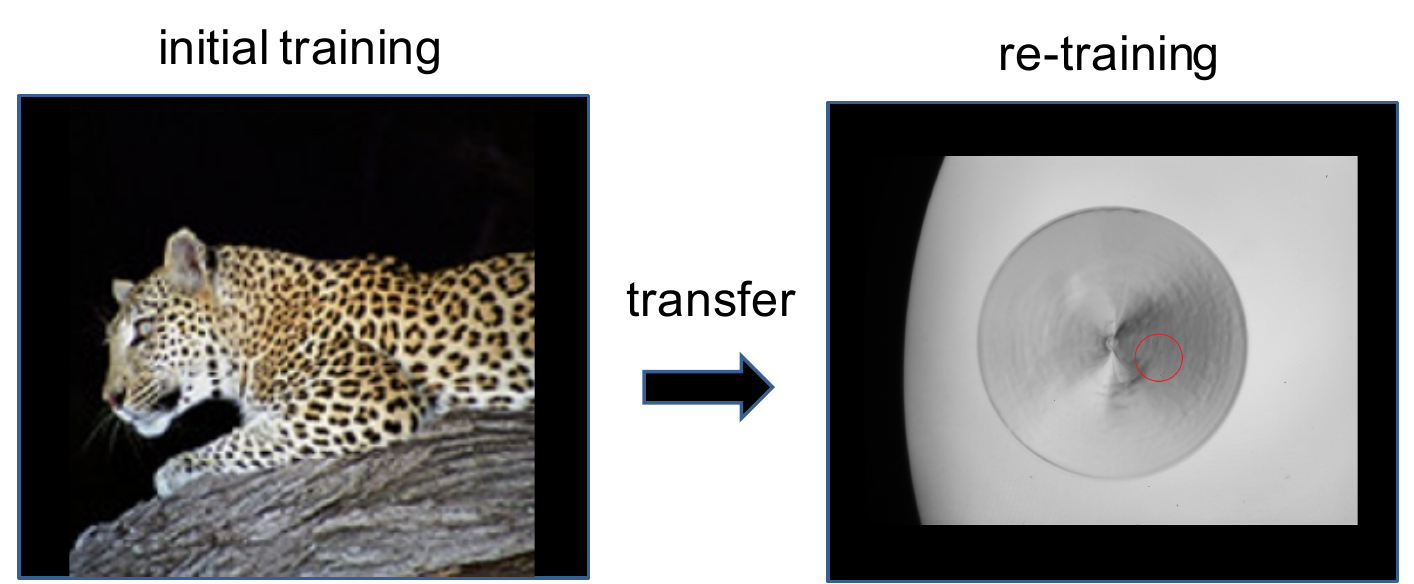}\caption{\label{fig:trans_example} Transfer learning is effective in scientific data applications.  Scientists at the National Ignition Facility, the world's largest laser, have used it to improve optical metrology for laser systems. After initial training on the ImageNet data set (sample image on left), the network was retrained on limited optics damage data (sample image on right) and was highly accurate at identifying defects.}
\end{figure}

While deep learning inherently relies on large data sets to train the many parameters in the network, it is also possible to develop networks using sparse data.  The key concept is called \emph{transfer learning} (see figure \ref{fig:transfer}).  In transfer learning, we first train a deep neural network on a large corpus of data.   This could be open source data, like ImageNet. Or, it might be scientific simulation data that is easier to obtain in large volumes than corresponding experimental observations.  In this initial training step, the network develops a representation for the data, developing an efficient latent space representation.  The model sets the full complement of parameters in this period.  If the task is image recognition, we might say that the network learns to see in this first step.  In the following step, a limited set of parameters, typically those in the last layer or layers of the network, are re-trained on a smaller corpus of data.  This data is typically more expensive data associated with a specialized task. Because only a limited number of parameters can be adjusted in the re-training step, we can get by with a much smaller data set.  Thus, transfer learning allows us to augment small, precious data sets with large, low-cost data sets to train effective networks.  This may sound too good to be true, but it works.  For example, scientists working at the National Ignition Facility trained a deep neural network classifier \cite{mundhenk} on ImageNet data (images of cats, fruits, etc.), but used subsequent transfer learning to help identify defects in high-power laser optics (images of damage sites in lenses) with greater than $95\%$ accuracy (figure\ref{fig:trans_example}).  Transfer learning potentially allows deep learning techniques to be applied to relatively small experimental data sets using augmentation from cheaper related simulation data sets or even unrelated open-source data sets.

\subsection{\label{sec:recurrent}Recurrent networks for time series}
\begin{figure}[h]
\centering{}\includegraphics[width=0.75\columnwidth]{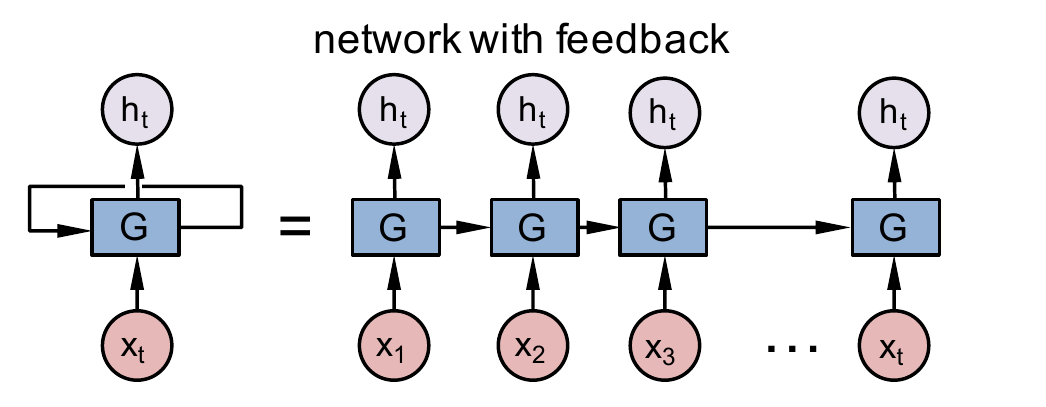}
\caption{\label{fig:rnn} Neural networks can be specialized to handle sequences of data.  Recurrent neural networks introduce feedback to deal with arbitrary length sequences. We show the recurrent network with an LSTM in an equivalent unfolded form.}
\end{figure}

We finally consider specializations for time series data.  The networks we have considered so far are feedforward networks.  Information that enters the network propagates through the network with each layer affecting only the subsequent layers.  However, when handling sequence information, like natural language or scientific time series, we may need to remind a layer of a value that it has seen before in the context of later values.  More specifically, we may want a feedback mechanism.  For this, we replace the simple neuron with a recurrent unit called a long short-term memory (LSTM) unit \cite{colah:rnn}.  The LSTM, more complicated than the feed forward neuron, uses feedback to establish a state of the unit.  Thus, the unit output is dependent not only on the current input from a sequence, but also on the state established by previous sequence values.  As shown in figure \ref{fig:rnn}, a recurrent network can be \emph{unfolded} to look like a feedforward network. The recurrent LSTM allows networks to adapt to sequences of arbitrary length and is a useful tool for analyzing records parameterized by time or other single scalar.

We summarize in table \ref{table:sum} the various networks and the tasks for which they might be appropriate.
\begin{table}[htpb]\small
\caption{\label{table:sum}Summary of network architectures and their uses to help with initial model selection.}
\begin{center}
\begin{tabular}{|>{\raggedright}p{0.8in}||>{\raggedright}p{1.1in}|>{\raggedright}p{1.1in}|>{\raggedright}p{1.1in}|>{\raggedright}p{1.1in}|>{\raggedright}p{1.1in}|}
\hline
\textbf{network type or technique} & fully-connected network & convolutional network & recurrent network & transfer learning & auto-encoder
\tabularnewline \hline
\textbf{application or data type} & scalar data & fixed-length vector or image data & time-histories & sparse data & data to be dimensionally reduced
\tabularnewline \hline
\textbf{learning category} & supervised & supervised & supervised & supervised & unsupervised
\tabularnewline \hline
\end{tabular}
\end{center}
\label{learning_deliverables}
\end{table}%

\section{\label{sec:computers}Impacts of machine learning on computer architectures}


Machine learning operations are readily parallelized.  This has made them amenable to execution on graphics cards with general-purpose GPUs, which are characterized by many-core processors and high memory bandwidth. Together with the CUDA language for writing arbitrary code on GPUs, numerous machine learning algorithms and software packages are taking advantage of this capability.  As practitioners looking to implement learning algorithms, we must choose the computer architecture for training carefully.  For the DJINN model \cite{humbird:djinn}, written in TensorFlow, training on a GPU proceeds about twice as fast as on an equivalent CPU. This puts competing design pressures on computers for scientific machine learning.  We may still want the good branching control, parallelism across large networks, and programming convenience of CPUs for scientific simulation.  For subsequent learning, we may want the benefits of GPUs for model training.  In some circumstances, machine learning workflows can benefit from specialized chips, sometimes called inference engines, used just to evaluate the already trained neural network.  Customers and computer vendors are increasingly considering heterogeneous architectures containing CPUs, GPUs, and inference engines.  However, the needs of computer users in the commercial technology, commercial goods, or scientific communities can be quite varied.  Our scientific community is responsible for exploring the computer design requirements generated by our research and developing a vision for the next generation of scientific computers.

\section{\label{sec:advancing}Jointly advancing physical science and machine learning}
Regardless of the particular task or the computer platform used, learning algorithms derive much of their power from their flexibility.  In fact, deep learning models achieve their tasks without detailed intervention by the user, say by explicitly constructing a parametric model.  Some go so far as to say that, for the most advanced algorithms, no one knows exactly how they function \cite{ai:dark_secret}.  Interpreting the function of these complicated algorithms is difficult, at least in part because there is often no external theory for the tasks they aim to achieve.  Their is no set of first principle laws for teaching autonomous vehicles or for parsing natural language text.  However, applied science is distinctly different.  For many tasks, like a regression task mapping numerical simulation inputs to their computed outputs, their exists at least an approximate parallel theory. Learned models for scientific tasks can be compared to a variety of existing theoretical models, they can be tested against repeatable experiments, and they can be checked against physical laws.  Moreover, the scientific community often produces its own data through simulation or experiment.  Thus, we can perform experiments on the learned models by augmenting or adapting training data with new examples to test the effects.

The use of modern machine learning for scientific purposes raises a long list of questions for exploration by the community.  Can we use machine learning to better understand experimental data?  Can we use machine learning to accelerate and improve numerical simulation?  How should we use learning to explore experimental design spaces?  How do we quantify uncertainty in analysis using machine learning?  Can we apply learning across data sets of multiple fidelities -- experiment, low-order simulations, higher-order simulations? Can we, as a scientific community, develop a more formal theory of machine learning by building on the foundations of statistical physics, for which there are many parallels?  With the proliferation of machine learning algorithms and software tools (table \ref{table:tools}) for implementing them, it is incumbent upon our community to embrace them and develop these tools to advance our scientific missions.

\begin{acknowledgments}
I would like to thank my Ensembles and Machine Learning Strategic Initiative team members for the challenging and exciting discussions that teach me so much.  Special thanks to Luc Peterson, John Field, Kelli Humbird, Jim Gaffney, Ryan Nora, Timo Bremer, Jay Thiagarajan, and Brian Van Essen. I also thank Jim Brase and Katie Lewis for inviting me into this research area and giving this kind of work an organized home at Lawrence Livermore National Laboratory. Prepared by LLNL under Contract DE-AC52-07NA27344.
\end{acknowledgments}

\begin{table}[htpb]\small
\caption{\label{table:tools}Tools and tutorials for getting started.}
\begin{center}
\begin{tabular}{|>{\raggedright}p{1.4in}|>{\raggedright}p{5.0in}|}
\hline
scikit-learn (Python) & \url{http://scikit-learn.org/stable/tutorial/basic/tutorial.html}\tabularnewline \hline
TensorFlow & \url{https://www.tensorflow.org/tutorials}
\tabularnewline \hline
Keras  & \url{https://keras.io/getting-started/sequential-model-guide}
\tabularnewline \hline
CNTK & \url{https://docs.microsoft.com/en-us/cognitive-toolkit/Tutorials}
\tabularnewline \hline
\end{tabular}
\end{center}
\end{table}%

\bibliographystyle{plain}
\bibliography{/Users/spears9/Files/presentations/conferences/APS_DPP_2017/machine_learing_tutorial/paper/spears_refs}
\end{document}